%% file: ecc18.tex
\documentclass[conference,10pt]{IEEEtran}

\usepackage{amssymb,amsmath,amsfonts,bm,epsfig,graphicx,theorem,latexsym}
\usepackage{rotating,setspace,latexsym,epsf,color,subfigure}
\usepackage{cite,authblk}

\newtheorem{Theorem}{Theorem}

\newtheorem{Lemma}{Lemma}

\newtheorem{assumption}{Assumption}

\input{./math_def}

\title{Optimizing Robustness against Cascading Failures under Max-Load Targeted Attack  \thanks{This
work is partially supported by the Department of Energy grant DE-OE0000779 and by the National Science Foundation grants CCF 1422165 and CPS 1646526.}}

\author[1]{Omur Ozel}
\author[2,3]{Bruno Sinopoli}
\author[3]{Osman Yagan\vspace{-0.05in}}
\affil[1]{\normalsize Department of Electrical and Computer Engineering, George Washington University, Washington, DC}
\affil[2]{\normalsize Department of Electrical and Systems Engineering, Washington University in St. Louis, MO}
\affil[3]{\normalsize Department of Electrical and Computer Engineering, Carnegie Mellon University, Pittsburgh, PA \vspace{-0.05in}}

\begin{document}

\IEEEoverridecommandlockouts
\maketitle

\begin{abstract}

Motivated by reliability of networks in critical infrastructures, we consider optimal robustness of a class of flow networks against a \textit{targeted} attack, namely max-load targeted attack, that triggers cascading failures due to removal of largest load carrying portion of lines. The setup involves a network of $N$ lines with initial loads $L_1, \ldots, L_N$, drawn from independent and identical uniform distribution, and redundancies $S_1, \ldots, S_N$ to be allocated. In the failure propagation mechanism, a line fails initially due to attack and later due to overloading. The load that was carried at the moment of failing gets redistributed equally among all remaining lines in the system. We analyze robustness of this network against the max-load targeted attack that removes the largest load carrying $p$-fraction of the lines from the system. The system designer allocates $S_i$ as a stochastic function of the load in each line. Assuming an average resource budget, we show that allocating all lines the free-spaces equally among nodes is optimal under some regulatory assumptions. We provide numerical results verifying that equal free-space allocation to all lines performs optimally in more general targeted attack scenarios.
\end{abstract}

\pagestyle{plain}
\pagenumbering{gobble}

\section{Introduction}

In a variety of networks, including power line networks, financial networks and transportation networks, dynamical interactions are observed in the form of dynamic flows. In flow networks, an event of at least one failure of a line can trigger other failures and, as a consequence, a significant portion of lines in the network may fail in the form of \textit{cascading failures}. Robustness of flow networks against cascading failures is an active research topic and has received interest from the academic community \cite{yagan2012optimal,HuangGaoBuldyrevHavlinStanley, Buldyrev,PahwaScoglioScala, Daniels1945, AndersenSornetteKwan}. The references \cite{PahwaScoglioScala, Daniels1945, AndersenSornetteKwan} consider power networks where the failure mechanism is the equal redistribution of load upon the failure of a power line. In this paper, we address a similar issue in flow networks and consider cascading failures in a {\em democratic fiber bundle}-like model. In particular, we build upon our recent work \cite{yagan_fiber_bundle,ozel2018robustness} and extend existing understanding on robustness of flow networks to the case of malicious \textit{targeted} attacks. This research also broadly contributes to the large body of work on information dissemination and influence propagation \cite{ TangYuanMaoLiChenDai,wu2017influence,YaganPRE, zhuang2016information,yagan2013conjoining}, percolation \cite{ParshaniBuldyrevHavlin,son2012percolation,radicchi2015percolation} and robustness \cite{li2012cascading,buldyrev2010catastrophic,gao2011robustness,brummitt2012suppressing}.  

In the problem setting, there are $N$ lines with initial loads $L_1, \ldots, L_N$ and free-spaces $S_1, \ldots, S_N$. The maximum flow allowed on a line $i$ is its {\em capacity}, and is given by $C_i=L_i+S_i$. When a line fails due to overloading, it is removed from the system and the load it was carrying (at the moment of failing) gets redistributed {\em equally} among all remaining lines in the system. We characterize the robustness of this network against a malicious \textit{targeted} attack that removes the largest load carrying $p$-fraction of the lines from the system. 

We consider a metric that measures the average number of lines that survive the cascade of failures over all possible attack sizes reminiscent of the one proposed in \cite{scheider2011}. In contrast to the previous work, current work focuses on the max-load targeted attack scenario where the adversary chooses the largest load carrying lines and removes them from the system initially. We extend our previous work and show under some regulatory conditions that among all allocations as ordered stochastic functions of initial loads robustness is maximized when all lines have the same redundancy regardless of their initial loads. This extension comes with pitfalls and challenges and we highlight and resolve them. In particular, we investigate the order statistics of independent and identically distributed random variables and extend the results \cite{YaganPRE,YaganNSR}, \cite{ozel2018robustness} to the max-load targeted attack.

In the rest of the paper, we explain the system model in detail, develop the problem and our framework in Section \ref{sec:Model}. We then characterize the optimal robustness using our framework in Section~\ref{sec:Optimal2}. In Section~\ref{sec:numerical} we show applications of the optimal robustness for targeted attacks. We conclude our paper in Section \ref{sec:Conc}. As notation: When we mention random variables (rvs), probability measure is denoted by $\mathbb{P}$, and corresponding expectation operator by $\mathbb{E}$. 

\section{The Network Model and the Problem}
\label{sec:Model}

\subsection{Network Model}

We consider a network of $N$ lines $\mathcal{L}_1, \ldots, \mathcal{L}_N$ with initial loads $\mathbf{L}=[L_1, \ldots, L_N]$ where $\mathbf{L}$ is an independent and identically distributed (i.i.d.) random sequence with uniform distribution over the interval $[L_{min},L_{max}]$ where $L_{{\sl min}} > 0$. We denote the average load $L_B = \mathbb{E}[L_i] = \frac{L_{min}+L_{max}}{2}$. The {\em capacity} $C_i$  of a line $\mathcal{L}_i$ defines the maximum power flow that can be carried by it, and is expressed as  
\begin{equation}
C_i= L_i + S_i, \qquad i=1,\ldots, N,
\label{eq:capacity}
\end{equation}
where $S_i$ denotes the {\em free-space} or {\em redundancy} allocated to line $\mathcal{L}_i$. We denote the vector of free-spaces as $\mathbf{S}=[S_1, \ldots, S_N]$. where $S_i \geq S_{min}$ for some $S_{min} \geq 0$. 

\subsection{Max-Load Targeted Attack and Initial Load Redistribution}

A {\em targeted} attack at time $t=0$, which we call {\em max-load targeted attack}, removes the largest load-carrying $p$-fraction of the lines. After the attack, the amount of load on each surviving line is given by its initial load plus its share of the total load of the failed lines. When a line fails due to an attack or its load exceeding its capacity, it is removed from the system and the load it was carrying (at the moment of failing) gets redistributed {\em equally} among all surviving lines. The load redistribution process continues recursively until no further failures occur, generating a {\em cascade of failures}. $n_{\infty}(p)$ denotes the {\em final} (i.e., steady-state) fraction of surviving lines when a $p$-fraction of lines are attacked initially. $n_{\infty}(p)$ is monotone decreasing in $p$. We are interested in the large graph asymptotic as $N \rightarrow \infty$. 

The increase in the initial load of every surviving line due to the initial attack is denoted by $Q_0(p) \geq 0$. Here, $Q_0(p)$ represents the asymptotic average extra load that has to be carried by surviving lines with respect to the randomness in $\mathbf{L}$. We denote the $k^{th}$ order statistic of the sequence $\{L_1, \ldots, L_N\}$ as $L_{(k:N)}$:
\begin{align}
L_{(1:N)} \leq L_{(2:N)} \leq \ldots \leq L_{(N:N)}
\end{align}
More precisely, the max-load targeted attack of size $p$ incurs an additional load to the surviving lines as:
\begin{align}\nonumber
Q^{(r)}_0(p) &\triangleq \lim_{N \rightarrow \infty} \frac{1}{N}\sum_{k=N-\lfloor pN \rfloor+1}^{N} L_{(k:N)} \\ &= \lim_{N \rightarrow \infty} \frac{1}{N}\sum_{k=N-\lfloor pN \rfloor+1}^{N}\mathbb{E}[L_{(k:N)}] \triangleq Q_0(p)\label{limit}
\end{align} 
where we assume $\lfloor pN \rfloor$ lines are attacked. Here, $Q^{(r)}_0(p)$ represents the random asymptotic time average extra load (is a random variable parametrized by $p$) and $Q_0(p)$ is its expected value (is a real number parametrized by $p$). We also denote $Q^{(r,N)}(p)\triangleq \frac{1}{N}\sum_{k=N-\lfloor pN \rfloor+1}^{N} L_{(k:N)}$ as the finite time average random extra load. Existence of the limit in (\ref{limit}) is established by using the law of large numbers for order statistics \cite[Theorem 3]{wellner1977glivenko} with an underlying uniform distribution of loads as well as using superadditivity of the sequence $\{\sum_{k=N - \lfloor pN \rfloor+1}^N\mathbb{E}[L_{(k:N)}]\}$ with $N$ and using Fekete's lemma \cite{steele1997probability}. Note that $Q_0(0)=0$ and $Q_0(p)$ is monotone increasing, bounded and continuous with
\begin{align}
Q_0(p) \leq Q_0(1) = L_B
\end{align}
We also define $F(p)\triangleq \frac{Q_0(p)}{1-p}$, which will be useful in the analysis. Note that $F(p)$ is monotone increasing since $Q_0(p)$ is monotone increasing. Note also that $F(p)=0$ at $p=0$ and $F(p) \rightarrow \infty$ for $p \rightarrow 1$.  As a result, $F$ is invertible over $s \in [0,\infty)$ with $F^{-1}(s)$ monotone increasing over $s \in [0,\infty)$.

\subsection{Calculating $Q_0(p)$ and $F(p)$}

Next, let us set $L_{min}=0$ and $L_{max}=1$ and assume that $L_i \sim U[0,1]$. In this case, it is well known that $L_{(k:N)} \sim Beta(k,N-k+1)$ where $Beta(r,s)$ represents the Beta distribution with parameters $r$ and $s$ with probability density function:
\begin{align}
p_{Beta}(x) = \frac{1}{B(r,s)}x^{r-1}(1-x)^{s-1}, \quad x \in (0,1)
\end{align}
where $B(r,s)=\frac{(r-1)!(s-1)!}{(r+s-1)!}$. The mean value of Beta distribution is known in closed form and we have:
\begin{align}
\mathbb{E}[L_{(k:N)}] = \frac{k}{N+1}
\end{align}
This enables us to obtain $Q_0(p)$ as
\begin{align}\label{eq:qp}
Q_0(p)=\lim_{N \rightarrow \infty} \frac{1}{N} \sum_{k=(1-p)N +1}^N \frac{k}{N+1} = p - \frac{p^2}{2}
\end{align}
Note that we dropped $\lfloor . \rfloor$ in the lower summation limit in (\ref{eq:qp}) as this causes no difference in the limit as $N \rightarrow \infty$. Consequently, we have
\begin{align}\label{abv}
F(p)=\frac{2p-p^2}{2-2p} =\frac{p}{2(1-p)} + \frac{p}{2}
\end{align}
We observe that $F(p)$ in (\ref{abv}) is convex and monotone increasing for all $p \in (0,1)$ as both $\frac{p}{1-p}$ and $p$ are monotone increasing and convex functions of $p$. By using proper scaling and shifting, we can generalize this observation for any uniform distribution over an arbitrary interval $[L_{min},L_{max}]$:
\begin{Lemma}
For uniformly distributed $L_i$, $F(p)=\frac{Q_0(p)}{1-p}$ is convex over $p \in (0,1)$. 
\end{Lemma}

\subsection{Failure Cascade Mechanism}

To understand the failure cascade mechanism, we will consider the fraction $f_t$ of lines that {\em fail} at time $t=0,1,\ldots$. The number of lines that are still alive at time $t$ is then given by $N_t = N (1-f_t)$ for all $t=0,1,\ldots$. The cascading failures start with an attack that targets a fraction $p$ of lines. Hence, we have $f_0 = p$. Upon the failure of these $f_0$ lines, their load will be redistributed to the remaining $(1-f_0)N$ lines, with each remaining line receiving an equal portion of the failed load. Since the attack of $pN$ lines causes $Q^{(r,N)}_0(p)N$ amount of load to be redistributed to the remaining lines, the resulting extra load per alive line is thus given by $\frac{Q^{(r,N)}_0(p)}{(1-f_0)}$. In the next stage, a line $i$ that survives the initial attack fails when its new load reaches its capacity:
\[
 L_i + \frac{Q_0^{(r,N)}(p)}{(1-f_0)} \geq L_i + S_i, 
\]
or, equivalently $S_i \leq \frac{Q_0^{(r,N)}(p)}{(1-f_0)N}$. Therefore, at stage $t=1$, an additional fraction of the lines that were alive at the end of the initial stage fail. This yields
\[
f_1= f_0 + \frac{\#\{i: \ S_i \leq \frac{Q_0^{(r,N)}(p)}{(1-f_0)}, \ L_i \leq L_{\lfloor pN \rfloor : N} \}}{(1-f_0)N}
\]
where $\#\{ \ \}$ notation is used to denote the cardinality of the set $\{ \ \}$. This process continues until there is no additional line failure and the stopping condition is valid in almost sure sense in view of the law of large numbers for order statistics. 

\subsection{Allocation of $S_i$ under Fixed Budget}

The objective of the system designer is to maximize the average number of surviving lines after the cascading failures in the network
\begin{align}\label{object}
J(\mathbf{S}) = \sum_{i=0}^{N} n_{\infty}(p_i)m_i
\end{align} 
for any network size $N$. Here, $p_i=\frac{i}{N}$ denotes the attack size, i.e., the ratio of attacked lines, and $m_i$ denotes the probability of $p_i$ on the support set $\{0,1/N, \ldots, (N-1)/N, 1\}$. The vector of probabilities is $\mathbf{m}=[m_0, \ldots, m_N]$. The defender wishes to maximize $J$ by selecting the vector $\mathbf{S}$. 
\begin{assumption} \label{asm1}
The probability of attack is monotone decreasing, i.e.:
\[
m_i \geq m_{i+1}, \quad \forall i
\]
\end{assumption}
This monotonicity reflects the fact that the attacker's resources can accommodate attacking larger number of lines with smaller probability. The system designer has a fixed budget for $S_i$:
\begin{align}
\frac{1}{N}\sum_{i=1}^N S_i \leq S_{B}
\end{align}
where $S_B$ is the average free-space budget available to the system designer. The allocation of $S_i$ is performed as a stochastic function of initial load $L_i$ only, and we denote $S_i(L_i)$ to reflect this stochastic dependence. This assumption is in line with typical network settings such as \cite{MotterLai,YaganPRE,YaganNSR,ozel2018robustness}. We make the following general assumption common to the referred network settings:
\begin{assumption} \label{asm2}
$S_i$ is stochastically ordered with respect to the initial loads $L_i$; that is,
\begin{align}
\mathbb{P}[S>s | L=\ell_1] \leq \mathbb{P}[S>s|L=\ell_2], \quad \forall s
\end{align}
whenever $\ell_1 \leq \ell_2$.
\end{assumption}
Note that this assumption encompasses the commonly accepted allocation $S_i=\alpha L_i$.  We solve the following robustness optimization problem under Assumptions \ref{asm1} and \ref{asm2}:
\begin{align} \nonumber
\max_{p_i(S_i|L_i)} & \ J(\mathbf{S})  \\ 
\mbox{s.t.} \ \ &\frac{1}{N}\sum_{i=1}^N S_i(L_i) \leq S_{B} \label{opt_problem}
\end{align}

\section{Optimal Robustness under the Max-Load Targeted Attack}
\label{sec:Optimal2}

We will show that distributing the available free-space $S_i$ as $S_i=S_B$ for all $i$ optimizes the robustness of the system under the specified load and free-space constraints. Note that since $S_i$ is a stochastic function of $L_i$, the sequence $\{S_i\}$ is i.i.d. as well. Next, in view of the fact that $S_i$ is stochastically ordered, $Q^{(r)}_0(p) \rightarrow Q_0(p)$ almost surely and cascading process may continue in the later stages, we have the following inequality 
\begin{align} \nonumber
    &n_{\infty}(p) \\ \nonumber &\hspace{-0.05in}\leq (1-p)\bP{S_{i} \geq \frac{Q^{(r,N)}_0(p)}{1-p} | L_i \notin \{ \textit{largest} \ \lfloor pN \rfloor \ L_i\}} \\  \nonumber &\leq (1-p)\bP{S_{k:N} \geq \frac{Q^{(r,N)}_0(p)}{1-p}, \forall k=1, \ldots, N-\lfloor pN \rfloor} \\ &\leq (1-p)\bP{S_i \geq \frac{Q^{(r)}_0(p)}{1-p}}
    \nonumber
    \\ & = (1-p)\bP{S \geq F(p)}  \label{rrr}
\end{align}
where $S_{k:N}$ denotes the $k$th order statistic of the sequence $\{S_1, \ldots, S_N\}$. In view of (\ref{rrr}), it suffices to work on a single random variable $S$ representing the sequence of free-space allocations. In particular, we wish to show that among all stochastic functions of $L_i$, distributing $\{S_i\}$ using the Dirac-delta distribution $\delta(s-S_B)$ is optimal. To this end, we first impose the condition that $S_i$ takes value from the set $\mathcal{S}=\{k \in \mathbb{N}: \ F(\frac{k}{N})\}$. Note that if $S_i=S_B$, then the final system size under this discrete Dirac-delta distribution is
\begin{equation}
n_{\delta,\infty}(p) = \left \{  
\begin{array}{cc}
1-\frac{i}{N}  &  \textrm{if $\frac{i}{N} < p_{\delta}^{\star} $}   \\
0  &     \textrm{if $\frac{i}{N} \geq p_{\delta}^{\star} $}
\end{array}
\right.
\label{eq:robustness_of_dirac_S}
\end{equation}
where the critical attack size $p_{\delta}^{\star}$ is given by
\[
p_{\delta}^{\star}=\lfloor F^{-1}(S_B) \rfloor.
\] 
In view of (\ref{eq:robustness_of_dirac_S}), the desired result will follow if we show
\begin{align}
  \sum_{i=0}^{N} n_{\infty}(p_i)m_i \leq  \sum_{i=0}^{Np_{\delta}^{\star}} (1-p_i)m_i
  \label{eq:to_show1}
\end{align}
We will get the desired result (\ref{eq:to_show1}) if we can show
\begin{align}\nonumber
    \sum_{i=0}^{N}  (1-p_i)\bP{S \geq F(p_i)}m_i \leq \sum_{i=0}^{Np_{\delta}^{\star}} (1-p_i)m_i,
\end{align}
or, equivalently that
\begin{align}\nonumber
    & \sum_{i=Np_{\delta}^{\star}+1}^{N}  (1-p_i)\bP{S \geq F(p_i)}m_i 
   \\
    &\quad \leq \sum_{i=0}^{Np_{\delta}^{\star}} (1-p_i) \bP{S < F(p_i)}m_i.
    \label{eq:to_show2}
\end{align}
Since $1-p$ is monotone decreasing over the range $0 \leq p \leq 1$, $m_i$ is monotone decreasing, and both $\bP{S \geq F(p)}$ and $\bP{S < F(p)}$ are non-negative, (\ref{eq:to_show2}) will follow if we can show
\begin{align}\nonumber
     \sum_{i=Np_{\delta}^{\star}+1}^{N} \bP{S \geq F(p_i)} 
    \leq \sum_{i=0}^{Np_{\delta}^{\star}}  \bP{S < F(p_i)},
\end{align}
or, equivalently that
\begin{align}
\sum_{i=0}^{N} \bP{S \geq F(p_i)} \leq \sum_{i=0}^{Np_{\delta}^{\star}} 1 = N p_{\delta}^{\star} + 1,
        \label{eq:to_show3}
\end{align}
Note that since $S_i$ is allowed to take value in $\mathcal{S}$, the left hand side of (\ref{eq:to_show3}) is equivalent to
\begin{align}
\frac{1}{N}\sum_{i=0}^{N} \bP{S \geq F(p_i)} = \int_0^1 \bP{S \geq F(p)} dp
\end{align}
This follows from the fact that $S$ is a discrete random variable with support over $\mathcal{S}$ and Riemann integral with any spacing of the form $\frac{1}{nN}$ with $n \in \mathbb{N}$ yields $\frac{1}{N}\sum_{i=0}^{N} \bP{S \geq F(p_i)}$.  

Next, we  make a change of variables $x=F(p)$ and write
\begin{align}\label{seq_2} 
    & \int_0^1 \bP{S \geq F(p)} \mathrm{d}p = \int_0^{\infty} \bP{S \geq x} \mathrm{d}\left(F^{-1}(x)\right) 
    \\ \nonumber & = \bP{S \geq x}F^{-1}(x) \bigr|_{x=0}^{\infty} - \int_0^{\infty} F^{-1}(x)\mathrm{d}(\bP{S \geq x}) \\ \label{seq_3} &= \int_{0}^{\infty} F^{-1}(x)\mathrm{d}(\bP{S \geq x}) \\& = \bE{F^{-1}(S)} \nonumber 
    \\ & \leq F^{-1}(S_B)  \label{eq:bound}
\end{align}
where we use integration by parts in (\ref{seq_2}), the fact that $S_i \geq F(p_c)$ for all $i$ in (\ref{seq_3}) and apply Jensen's inequality for the function $F^{-1}(x)$ over the interval $x \in (0,\infty)$ to reach (\ref{eq:bound}) since it is concave in $x$ over this interval. Combining the results, we conclude that
\begin{align}
\sum_{i=0}^{N} \bP{S \geq F(p_i)} \leq N F^{-1}(S_B) &= N p_{\delta}^{\star} \\ &\leq N p_{\delta}^{\star} + 1
\end{align}
This establishes (\ref{eq:to_show3}) and the desired result (\ref{eq:to_show1}) follows in view of the preceding arguments. We established the following:
\begin{Theorem}
For uniformly distributed $L_i$, $S^{*}_i=S_B$ for all $i$ is optimal for (\ref{opt_problem}).
\end{Theorem}

Note that the solution in above theorem bears the meaning that irrespective of the initial load distribution, identical allocation of the free-space to all lines is optimal. In other words, the robustness is maximized by choosing a line's capacity $C_i$ through $C_i = L_i + S_B$ no matter what its load $L_i$ is.

\subsection{General Load Distributions and Bounds on Order Statistics}

\begin{figure}[!t]
\centering{
\hspace{-0.5cm} 
\includegraphics[totalheight=0.28\textheight]{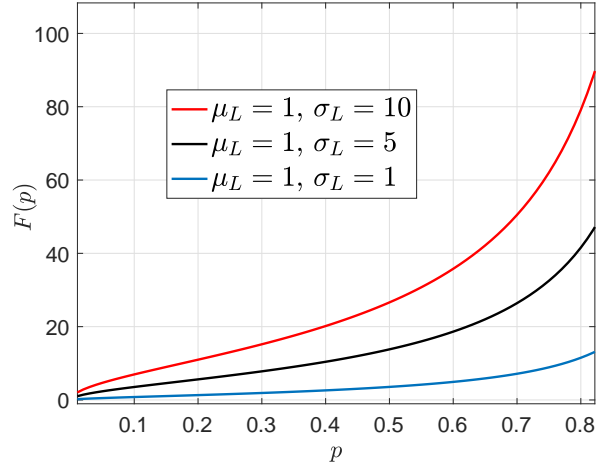}\vspace{-0.1in}}
\caption{\sl We plot $F(p)$ resulting from the upper bound in (\ref{upper}) for different $\sigma_L$ and fixed $\mu_L$.}\vspace{-0.3in}
\label{fig:2} 
\end{figure}

In general, deriving analytic results to work with the order statistics is not an easy task. In order to calculate $\mathbb{E}[L_{(k:N)}]$, one has to compute the following integral:
\begin{align}\nonumber
\mathbb{E}[&L_{(k:N)}] \\ &\  = \int_0^{\infty} x p_L(x) {N-1 \choose k-1} P_L(x)^{k-1}(1-PL(x))^{N-k} dx \label{exact}
\end{align}
and closed form expressions for this integral are not in general available. Moreover, as the network size $N$ grows, the term ${N-1 \choose k-1}$ becomes hard to obtain let alone the whole integral. We, therefore, observe that it requires significant numerical effort to obtain $Q_0(p)$ especially when the network size $N$ is large and the distribution of $L_i$ is arbitrary. These issues add to the difficulty of generalizing law of large numbers for arbitrary distributions. In order to obtain a more tractable analysis, we next use the upper bound for $\mathbb{E}[L_{(k:N)}]$ from \cite[Eq. (2)]{bertsimas2006tight}, which is attributed originally to reference \cite{arnold1979bounds} and known to be tight for some distributions
\begin{align}\label{upper}
\mathbb{E}[L_{(k:N)}] \leq \mu_L + \sigma_L \sqrt{\frac{k-1}{n-k+1}}
\end{align}
where $\mu_L$ is the mean and $\sigma_L$ is the variance of the distribution of $L_i$. We approximate $\mathbb{E}[L_{(k:N)}]$ using the bound in (\ref{upper}). In Fig. \ref{fig:2}, we plot the function $F(p)$ resulting from using the upper bound in (\ref{upper}) in the calculation of $Q_0(p)$ in (\ref{limit}). It is observed numerically that increasing the mean value $\mu_L$ decreases the critical value $p^{c}$ for which the function $F(p)$ turns convex. In contrast, increasing the value of the standard deviation $\sigma_L$ increases $p^{c}$. Still, in our numerical results we observe that the value of $p^{c}$ remains minor with respect to the mean $\mu_L$ especially when the variance $\sigma^2_L$ is moderately small. Hence assuming the convex portion of $F(p)$ for analysis does not yield restriction. We also corroborate this behavior through extensive numerical results using the exact form in (\ref{exact}) for various distributions. 


\subsection{Recovering Optimal Robustness under Random Attacks}

In our recent work \cite{ozel2018robustness}, we consider random attacks and we show that under a statistical distribution on loads $L_i$ and free-spaces $S_i$, a Dirac delta distribution at $\bE{S}$ is optimal irrespective of the distribution of $L$, generalizing the approaches in \cite{YaganPRE,YaganNSR} for the general robustness metric $\int_0^1 n_{\infty}(p)dp$. The framework in the current paper is a generalization of \cite{ozel2018robustness} for the max-load targeted attack. In order to recover the previous results, let us assume that the attacker ``randomly" removes a $p_i=\frac{i}{N}$ fraction of lines (i.e., removes all possible $i$ lines with equally likely probability) from the system. Then, the extra load incurred to the remaining $1-p_i$ portion of lines is $Q_0(p_i)=p_i L_B$. In this case, the function $F(p)$ is
\begin{align} \label{fcn1}
F(p) = \frac{p L_B}{1-p}
\end{align}
We observe that $F(p)$ in (\ref{fcn1}) is strictly convex for all $p \in (0,1)$. Following the steps in Theorem 1, we can guarantee that distributing $S_i=S_B$ for all $i$ maximizes robustness in (\ref{object}). Our analysis also guarantees that convergence observed in the mean-field analysis in \cite{ozel2018robustness} is in almost sure sense.

\section{Numerical Results}
\label{sec:numerical}

The results presented in the paper are based on a model with a mean field assumption in that when a line fails, its load gets redistributed globally and equally among all active lines in the network. This mean field assumption may not hold for networks with possible local redistribution behavior such as power networks. On the other hand, models based on exclusive local redistribution cannot capture the long-range nature of physical laws. Consequently, a model where the failed load is redistributed both locally and globally would be more suitable. In addition, we assume  a uniform distribution of loads in order to be able to invoke law of large numbers for order statistics. In this section, we perform simulations under a topology-based redistribution model with loads having distributions other than uniform. We wish to test how the derived optimal scheme of equal free-space distribution holds when the load associated with failed nodes is redistributed (at least in part) locally according to a network topology under different load distributions. 

In the spirit of the redistribution models presented in \cite{MotterLai, Mirzasoleiman}, we consider a topology based redistribution model that combines both local and global redistribution behaviors. This extended model gauges the locality of the redistribution by the parameter $\gamma \in [0,1]$ and the network topology is generated as an Erd\H{o}s-R\'enyi graph $\mathbb{G}(n,N)$. At each stage, the portion of the load that is not absorbed from each failed line is divided into two parts: $\gamma$-fraction is redistributed locally among neighboring lines (with each neighbor receiving an equal portion), and $(1-\gamma)$-fraction is redistributed equally among {\em all}  surviving lines (irrespective of topology). In this model, setting $\gamma=0$ recovers the mean-field model introduced in Section \ref{sec:Model}, while setting $\gamma=1$ gives a merely topology based redistribution model. We compare the robustness of this system with respect to the following metric: 
\begin{align} \label{eq:robmetric}
    \mathcal{R} = \frac{1}{N}\sum_{i=1}^{N}n_{\infty}(p_i)
\end{align}
where $p_i=\frac{i}{N}$. We run simulations for loads with Weibull distributions with fixed $\bE{L}$ and $\bE{S}$ while varying the scale parameter $k$ of the distribution and compare robustness with respect to $\mathcal{R}$. 

\begin{figure}[t]
\centering{
\hspace{-0.15cm} 
\includegraphics[totalheight=0.28\textheight]{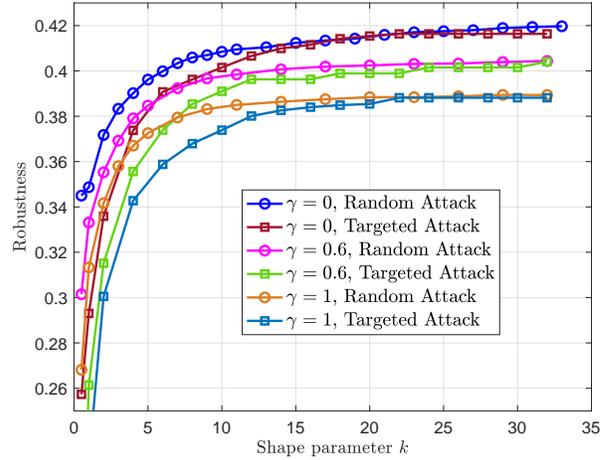}}\vspace{-0.05in}
\caption{\sl We let $L_1, \ldots, L_N$ be obtained from Weibull distribution with $L_{\textrm{min}}=1$ and $k,\lambda$ such that $\bE{L}=2$ and $S=1.74L$. For $\gamma=0$, $\gamma=0.6$, and $\gamma=1.0$, the robustness $\mathcal{R}$ is plotted with respect to the shape parameter $k$.}\vspace{-0.15in}
\label{fig:local1} 
\end{figure}

For this new redistribution model, we perform simulations where we set the number of nodes to $n=250$ and number of edges as $N=8400$, and create a random network according to the Erd\H{o}s-R\'enyi $\mathbb{G}(n,N)$ model. This leads to each line having on average $132$ neighbors. The loads $\{L_1,\ldots,L_N\}$ are drawn from i.i.d. Weibull distribution (with $L_{\textrm{min}}=1$ and $k,\lambda$ such that $\bE{L}=2$) and the free spaces are set as $S_i=1.74L_i$. We run $100$ independent experiments for each parameter set, and report the average value of the robustness metric defined in (\ref{eq:robmetric}). The results of simulations are shown in Fig. \ref{fig:local1} for $\gamma=0$, $\gamma=0.6$ and $\gamma=1$. We observe that $\mathcal{R}$ is maximized in the limiting case $k \rightarrow \infty$. This observation points to the optimality of Dirac delta distribution since the Weibull distribution approaches the Dirac delta function as $k$ grows large. These numerical findings show that the uniform allocation among the loads, in the case of the max-load targeted attack under Weibull distributed loads with possible (partial) {\em local} load redistribution, performs well even when the redistribution of the loads on failing lines are performed based on the topology of the network. Note also that increasing the locality parameter $\gamma$ decreases robustness, which is in line with our earlier findings in \cite{ozel2018robustness} for randomly generated attacks.

\begin{figure}[t]
\centering{
\hspace{-0.15cm} 
\includegraphics[totalheight=0.28\textheight]{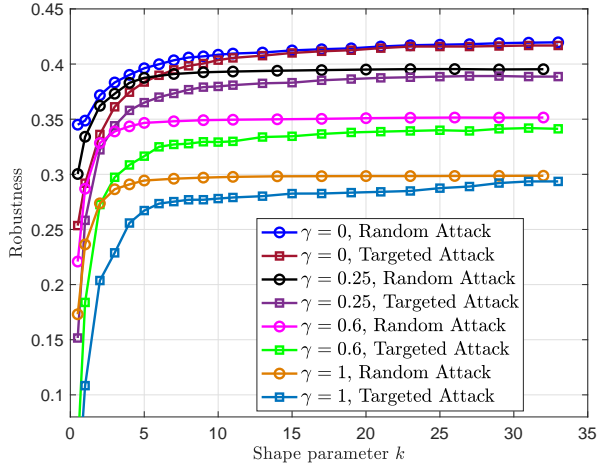}}\vspace{-0.05in}
\caption{\sl We repeat the experiments in Fig. \ref{fig:local1} for $\gamma=0.25$, $\gamma=0.6$, and (c) $\gamma=1.0$, under a different Erd\H{o}s-R\'enyi graph with significantly smaller average number of neighbors and provide comparison of random and targeted attack scenarios. We also include the comparison for $\gamma=0$ case.}\vspace{-0.1in}
\label{fig:local2} 
\end{figure}

We finally test the effect of node degree on the robustness of the network. We consider the Erd\H{o}s-R\'enyi $\mathbb{G}(n,N)$ model with the same number of nodes $n=250$ and a smaller number of edges $N=625$. In this case, each line has $8$ neighbors on average. The results are presented in Fig. \ref{fig:local2} for $\gamma=0$, $\gamma=0.25$, $\gamma=0.6$ and $\gamma=1$, respectively. We observe that the robustness of the network improves irrespective of locality parameter $\gamma$. Additionally, when compared to the results in Fig. \ref{fig:local1}, we observe that robustness is affected by the connectivity of the network in different ways for different $\gamma$. In particular, the drop in robustness with $\gamma$ is more significant for a loosely connected network. 

\section{Conclusions and Future Work}
\label{sec:Conc}

In this paper, we investigated the optimal robustness for a class of flow networks against cascading failures under max-load targeted attacks that remove the largest load carrying $p$-fraction of the lines. In the following stages of the failure cascade mechanism, a line is removed due to overloading. The load that was carried at the moment of failing gets redistributed equally among all remaining lines in the system while the rest of the load is lost. Assuming an average resource constraint on free-space $S_i$, we show that among all possible allocations as ordered stochastic functions of loads the optimum is achieved when all lines are allocated the same free-spaces under some conditions. 

Our current results reveal that the methodology developed in \cite{ozel2018robustness} can be extended to deal with max-load targeted attacks. In future work, we will focus on understanding  the critical attack size necessary to disrupt the whole network and the resulting tri-critical behavior. We will also extend our framework to understand the attacker-defender interactions in single and multiple stages of the cascade formation in multiple interdependent networks by devising necessary cost structures.

\end{document}

%% file: math_def.tex
\def \n2{{N_0 \over 2}}

\newcommand{\bP}[1]{{\mathbb{P}}\left[{#1}\right]}
\newcommand{\bE}[1]{{\mathbb{E}}\left[{#1}\right]}

\def \h5{\hspace{0.5in}}